\documentclass[aps,pre,twocolumn,amsmath,amssymb,superscriptaddress,showpacs,longbibliography]{revtex4-1}

\usepackage{comment}
\usepackage{float}
\usepackage{color}
\usepackage[usenames,dvipsnames,svgnames,table]{xcolor}
\usepackage{graphicx}
\usepackage{dcolumn}
\usepackage{bm}

\usepackage{soul}

\begin{document}
\title{Metastable localization of diseases in complex networks 
} 
\author{R. S. Ferreira}
\affiliation{Departamento de Ci\^{e}ncias Exatas e Aplicadas, Universidade Federal de Ouro Preto, 35931-008 Jo\~{a}o Monlevade, Brazil}

\author{R. A. da Costa}
\affiliation{Department of Physics \& I3N, University of Aveiro, 3810-193 
Aveiro, Portugal
}

\author{S. N. Dorogovtsev}
\affiliation{Department of Physics \& I3N, University of Aveiro, 3810-193 
Aveiro, Portugal
}
\affiliation{A.~F. Ioffe Physico-Technical Institute, 194021 St.~Petersburg, Russia
}

\author{J.  F.  F. Mendes}
\affiliation{Department of Physics \& I3N, University of Aveiro, 3810-193 
Aveiro, Portugal
}

\begin{abstract}
We 
describe the phenomenon of localization 
in the epidemic SIS model on highly heterogeneous networks in which strongly connected nodes (hubs) play the role of centers of localization. 
We find that in this model the localized states below the epidemic threshold are metastable. 
The longevity and scale of the metastable outbreaks do not show a sharp localization transition, instead there is a smooth crossover from 
localized to 
delocalized 
states as we approach the epidemic threshold from below.  
Analyzing these long-lasting local outbreaks 
for 
a random regular graph 
with a hub, we show how this localization can be detected from the shape of the distribution of the number of infective nodes. 
\end{abstract}

\maketitle

\section{Introduction}

Localization is one of the key phenomena in nature. It was extensively explored in a wide range of systems including localization of electrons in disordered systems, localization of phonons, and many others \cite{anderson1958absence,lifshitz1964energy,
thouless1974electrons,kirkpatrick1985localization,jahnke2008wave,
odor2014localization,pastor2015distinct}. Recently it attracted much attention in application to 
epidemic spreading \cite{goltsev2012localization,lee2013epidemic}, where localization means persistence of an island of disease below the epidemic threshold around a strongly connected node or a dense cluster in a network. 
The 
complication is that the SIS (susceptible-infective-susceptible) epidemic model 
has an absorbing state in which infection is absent \cite{anderson1991infectious,cardy1985epidemic,pastor2015epidemic}, and so below the epidemic threshold islands of disease with a finite number of infective nodes cannot survive forever. In other words, a system with a finite number of infected nodes has a non-zero probability to recover immediately. For a large but finite number of infected nodes, however, this probability is small, so the complete recovery can take a long time. For finite fully connected graphs, this behavior was described in Ref.~\cite{deroulers2004field}. Consequently, in the heterogeneous SIS model, localization should be only metastable \cite{sahneh2016delocalized}, manifesting itself in the form of long-lasting local outbreaks of the disease below the epidemic threshold.

In this paper we consider the SIS model on a random regular graph 
with a single hub (``spreader center'') and investigate the metastable nature of localization of a disease. 
 On a regular network, the SIS model is equivalent to the ordinary contact process \cite{cardy1985epidemic} that belongs to the directed percolation universality class \cite{hinrichsen2000non,odor2004universality,henkel2008non}. These processes can be solved exactly only on fully connected graphs, so we have to resort to extensive numerical simulations. 
On the other hand, the heterogeneous (annealed) \cite{pastor2001epidemic,castellano2014theoretical} and quenched \cite{van2011n,van2012epidemic,
goltsev2012localization,castellano2014theoretical} mean-field approaches do not take into account the fluctuations and the absorbing state in the SIS model, so they cannot provide even a qualitative description of metastable localization.  
In the present paper we show how the effect of this localization can be detected by analyzing the shape of 
the distribution of the number of infected nodes in the metastable state. 
By metastable state we mean the active quasistationary state below the epidemic threshold. 
Measuring the lifetime of the localized states 
we describe how the metastable localization depends on the epidemic parameter $\lambda$. 
From the distribution of the number of infective nodes in finite graphs we extract the contribution of the metastable localized states and compare it with the solution of the SIS model on a star graph, uncovering the effect of the network substrate. 
We observe two regimes, localized and delocalized, separated by a smooth crossover occurring in a region around $\lambda_{\text{crossover}}{<}\lambda_c$. Surprisingly, in contrast to predictions of the quenched mean-field theory, 
the disease is localized on a hub below the crossover region, and between $\lambda_{\text{crossover}}$ and $\lambda_c$, the effect of hub disappears.

\section{Model}

We study the 
SIS model on a random regular graph of $N$ nodes, with all but one having degree $k$. The single hub 
in this graph is a node with $q\gg k$ connections. For the sake of comparison we also consider the same network without hub ($q=k$). The graph has finite length loops (cycles) only when $N$ is finite. 

In the SIS model, each node can be in two states: susceptible and infective. 
A susceptible node is infected by each of its infective nearest neighbors with
rate $\lambda$ (so-called effective spreading rate), which is the control parameter in this model. An infective node spontaneously recovers with unit rate. 
Our simulations  
for uniform networks up to $2\times 10^9$ nodes confirmed that in the infinite random regular graph with the coordination number $k=6$, the epidemic threshold is $\lambda_c = 0.2026(1)$ Ref.~\cite{mata2013pair}. This is the usual continuous transition in the contact process above the upper critical dimension \cite{cardy1985epidemic,hinrichsen2000non,odor2004universality,henkel2008non,dorogovtsev2008critical, ferreira2013critical,cai2016solving}.

\section{Simulations}

The SIS dynamics is 
implemented as follows~\cite{boguna2013nature,ferreira2012epidemic}. 
During the process we trace the numbers of infective nodes, $n(t)$, and of active links, $\ell(t)$. 
By definition, an active link is a link of an infective node, and the links between two infective nodes are counted twice in $\ell(t)$.  
At each step, with probability $n(t)/[n(t)+\lambda \ell(t)]$ a uniformly randomly chosen infective node becomes susceptible. 
With complementary probability $\lambda \ell(t)/[n(t)+\lambda \ell(t)]$ an active link is chosen uniformly at random. If it connects infective and susceptible nodes, then the susceptible one becomes infective. If both nodes are infective, nothing occurs. Finally, time is incremented by $1/[n(t)+\lambda \ell(t)]$.

%
\begin{figure}[t]
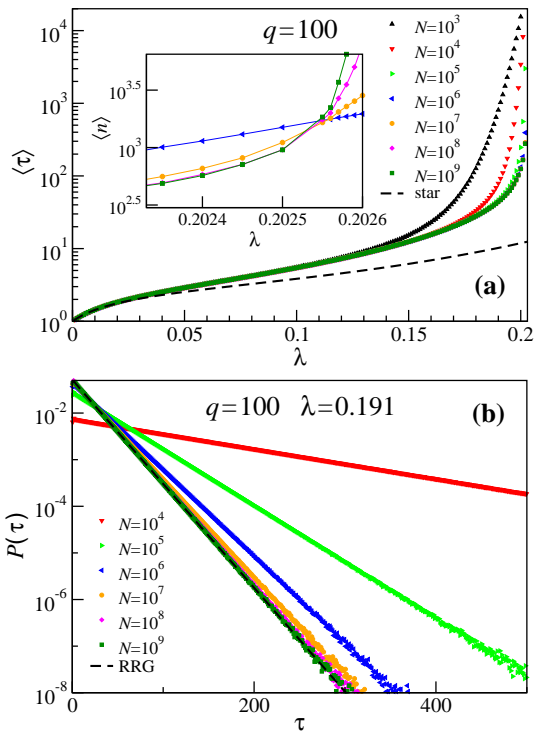

\centering
\hspace{1pt} \includegraphics*[width=6.97cm]{figure1a}
\\[3pt]
\includegraphics*[width=7cm]{figure1b}
\caption{Time to reach the absorbing state of the process for the random regular networks, $k=6$, with a single hub of degree $q=100$ below the epidemic threshold $\lambda_c=0.2026(1)$. (a) Average lifetime of the process vs. $\lambda$ when each realization starts from the state in which only the hub is infective (spreading experiment). 
For small $\lambda$, the curves approach the dependence $\langle \tau \rangle(\lambda)$ obtained from Eq.~(\ref{ee50}) for the corresponding star graph (dashed curve). 
The inset shows the average number of infective nodes in the spreading experiment near the epidemic threshold. 
(b) The distribution of times between the attempts to fall into the absorbing state in quasistationary simulations at $\lambda=0.191<\lambda_c$. The dashed line stands for the uniform network (random regular graph), $q=k=6$.
}
 \label{fig:spreading}
 \end{figure}
%

We perform numerical simulations in two ways. 
In the first approach, which we call spreading experiment, in each realization, the hub is initially infective and the other nodes are susceptible, and the process finishes when it reaches the absorbing state (all nodes susceptible). 
In the second approach, we obtain the quasistationary distribution of active nodes by the method of Refs.~\cite{ferreira2011quasistationary2,oliveira2006quasi,de2005simulate}
that excludes the absorbing state from the simulations. 
When the SIS process reaches the absorbing state, we restore one of the previous active configuration taken at random from the history of the process. 
This procedure optimizes the numerical simulations confining the dynamics of the process to active states, which enables us to efficiently collect the statistics of the quasistationary regime independently on initial conditions. 
For more details about our simulation method see the Appendix. 

Figure~\ref{fig:spreading} presents the statistics for the lifetime of the process obtained by implementing these two approaches. In Fig.~\ref{fig:spreading}(a) we show the average time to reach the absorbing state (average lifetime) in the spreading experiment versus the parameter $\lambda$ for different network sizes $N$. In the infinite network, the lifetime diverges at the epidemic threshold. For $\lambda$ sufficiently small, all curves for different $N$ collapse into one. As $\lambda$ increases, these curves separate from each other 
due to the loops of a finite length which are present in the finite networks. 
These loops increase the average lifetime due to the additional reinfection of the hub occurring when disease spreads through a loop. In the locally tree-like infinite networks, loops are infinite, and reinfection is possible by only returning to the hub through the same path. The effect of loops is stronger in the small networks, and the curves start to separate at smaller 
$\lambda$ as $N$ decreases. 
The inset of Fig.~{\ref{fig:spreading}}(a) demonstrates a strong size effect on the average number of infective nodes $\langle n \rangle$ near the epidemic threshold. The average $\langle n \rangle$ in Fig.~{\ref{fig:spreading}}(a) is taken over the entire time of the spreading experiment and over realizations.
In the quasistationary simulations, the lifetime of the process can be 
extracted from the 
distribution of times between the attempts to reach the absorbing state,  
see Fig.~\ref{fig:spreading}(b) for $\lambda =0.191<\lambda_c$. 
The figure shows that this distribution approaches the exponential distribution for a random regular graph as the network size goes to infinity.

\begin{figure}[t]
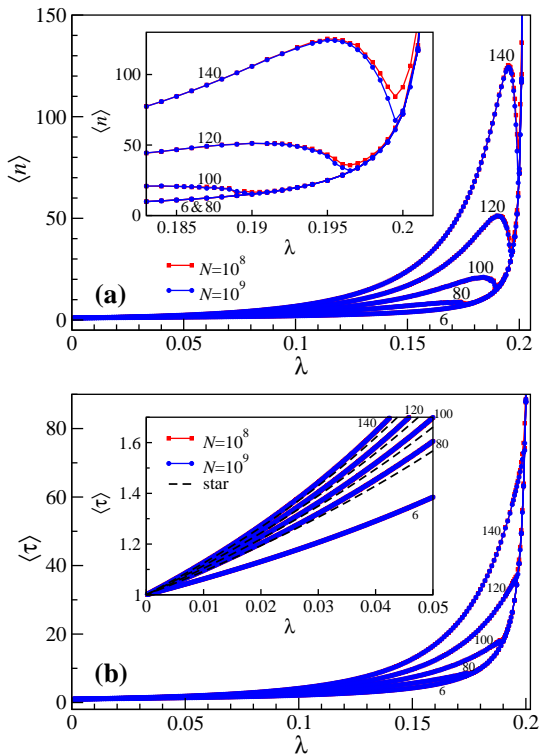

\centering
\hspace{-9.5pt} \includegraphics*[width=7cm]{figure2a}
\\[5pt]
\includegraphics*[width=6.95cm]{figure2b}
\caption{(a) Average number of infective nodes $\langle n \rangle$ and (b) the average lifetime $\langle \tau \rangle$ in the quasistationary regime of the SIS model on
random regular graphs with 
a single hub 
vs. $\lambda 
<\lambda_c\approx 0.2026$. 
The coordination number of the graph is $6$. 
The curves are for two network sizes, $10^8$ and $10^9$ nodes, and different values of the hub degree $q$ (see the numbers on the curves). 
The inset of panel (a) shows separation of 
the curves for networks of $10^8$ and $10^9$ nodes 
in the crossover region for each given $q$.  
The inset of panel (b) shows the region of small $\lambda$ in which the curves are close to the 
solutions of Eq.~(\ref{ee50}) (dashed lines) for the corresponding star graphs with $q$ leaves.
}
\label{fig:nVSkmax}
\end{figure}

Figure~\ref{fig:nVSkmax} shows how the average number of infective nodes $\langle n \rangle$ and the average lifetime $\langle\tau\rangle$ in the quasistationary regime depend on $\lambda$ for different hub degrees and network sizes.    
We observe a strong dependence of $\langle n \rangle$ and $\langle\tau\rangle$ on $q$ for $\lambda$ below $\lambda_{\text{crossover}}(N,q)$. In this range of $\lambda$, $\langle n \rangle$ and $\langle\tau\rangle$ rapidly grow with $q$, which indicates that the disease is localized around a hub and survives for much longer times than in the homogeneous network. 
Above $\lambda_{\text{crossover}}$, the effect of the hub disappears, and the curves $\langle n \rangle(\lambda)$ and $\langle\tau\rangle(\lambda)$ collapse to the ones for the uniform random regular graph. The inset of Fig.~\ref{fig:nVSkmax}(a) demonstrates that there is not a sharp transition from the localized state to delocalization but rather a smooth crossover between the two regimes. The dependence of $\langle n \rangle$ on the network size is well seen only in the region of this crossover. Notably, the curves $\langle n \rangle(\lambda)$ show a pronounced peak near $\lambda_{\text{crossover}}$ in contrast to $\langle\tau\rangle(\lambda)$. Finally, the inset of Fig.~\ref{fig:nVSkmax}(b) depicts $\langle\tau\rangle$ for small $\lambda$, where the results of the simulations are close to the lifetimes of the corresponding star graphs with $q$ leaves.

%
%

To quantify the effect of the hub we consider 
the complete quasistationary distribution of the number of infective nodes, $p_n$. This distribution is the probability that at a random instant an active system contains $n$ infective nodes. 
Figure~\ref{fig:ff1}(a),(c),(e) shows the distributions $p_n$ for different hub degrees $q$. For each $q$, we choose the epidemic parameter $\lambda$ in the corresponding crossover region, and measure $p_n$ for different network sizes. For the sake of comparison, we show $p_n$ for the random regular graph. 
The insets in these panels demonstrate a significant separation of these curves at large $n$ (log-linear plots). 
To validate our results, we checked that $p_1$ perfectly agrees with the inverse first moment of the distribution    
$P(\tau)$, which is a fundamental relation 
\cite{dickman2002quasi}.

\section{Decomposition of the distribution}

The effect of a hub is local, which results in the contribution of the order of $1/N$ to full distribution $p_n$ for large networks. Let us extract this contribution from the measured distribution $p_n(\lambda,q,N)$ for networks of different sizes by assuming the following ansatz: 
\begin{equation}
p_n=A(N)\left[p_{n,k}+\frac{H_n}{N}\right],
\label{ee10}
\end{equation}
where $p_{n,k}$ is the distribution of the number of
infective nodes in the uniform random regular networks 
with coordination number $k$, $H_n$ is a yet unknown function, and 
$A(N) = [1+\sum_n H_n/N]^{-1}$ 
is a normalization 
factor. 
In this ansatz, 
$H_n$ is independent of $N$ in the limit of large $N$ and is determined only
by $\lambda$, $k$, and $q$. 
We first assume the form of Eq.~(\ref{ee10}) and then we shall validate it analyzing results of our simulations. The rationale behind this form is the following. As the network size goes to infinity, the region where activation of infective nodes is influenced by the hub remains finite. This region is the same as for the Bethe lattice with a hub. On the other hand, far away from the hub, active states in the quasistationary regime are similar to those for the uniform random regular graph, this area grows as $N$, and so the activity occurs mostly far from the hub. Consequently one can expect that the the relative contribution of the localized states for the total distribution $p_n$ indeed scales as $1/N$.  

%
\begin{figure}[t]
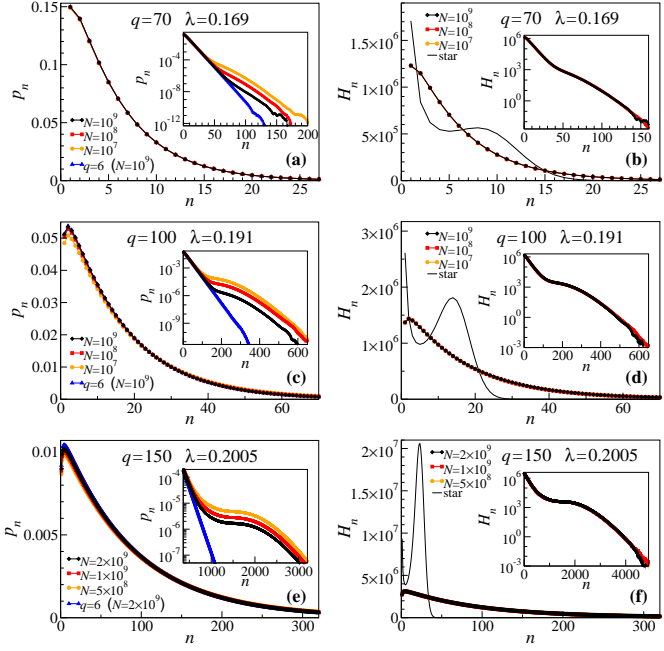

\includegraphics*[width=117pt]{figure3a}\ \  \includegraphics*[width=122.9pt]{figure3b}
\\[4pt]
\hspace{-0.2pt}\includegraphics*[width=117.3pt]{figure3c}\ \  \includegraphics*[width=122.9pt]{figure3d}
\\[4pt]
\hspace{-6pt} \includegraphics*[width=119.9pt]{figure3e}\ \  \includegraphics*[width=122.9pt]{figure3f}
\caption{
(a), (c), and (e) Distribution $p_n$ of the number of infective nodes in the quasistationary regime for different hub degree $q$ and values of $\lambda$ in the crossover region. Three different network sizes are presented in each panel. 
These plots also show the curves for uniform random regular graphs $q=k=6$. The insets demonstrate the separation of these curves in log-linear representation. 
(b), (d), and (f) Curves $H_n$ extracted from $p_n$ 
by employing Eq.~(\ref{ee30}). 
The initial values $H_1=1.2\times10^6,\ 1.4\times 10^6$ and $2.7\times 10^6$ in (b), (d), and (f), respectively, enable the collapse of the curves for different network sizes into one in each of the panels. 
The thin black lines show the stationary distributions $p_n^{\text{star}}$ for the corresponding star graphs  multiplied by $\sum_n H_n$  for the sake of comparison. The insets demonstrate the collapse of the curves of the main panels in log-linear representation. 
}
\label{fig:ff1}
\end{figure}

The direct 
application of Eq.~(\ref{ee10}) to extracting the function $H_n$ from the numerical data
obtained in our simulations for $p_{n}$ and
$p_{n,k}$ requires the knowledge of the
normalization factor $A(N)$.
We remove this unknown factor from the calculations by
rewriting Eq.~(\ref{ee10}) as
\begin{equation}
 \frac{p_{n+1}}{p_n} = \frac{p_{n+1,k}+H_{n+1}/N}{p_{n,k}+H_n/N}.
\label{ee20}
\end{equation}
Using this equation we express 
$H_{n+1}$ in terms of $H_n$ and the
distributions $p_n$ and $p_{n,k}$,
\begin{equation}
H_{n+1}=[Np_{n,k}+H_n]\frac{p_{n+1}}{p_n}-Np_{n+1,k}.
\label{ee30}
\end{equation}
For a given $\lambda$ and an arbitrary initial value $H_1$, we can extract the function $H_n$ by iteratively applying this recursive equation to the numerical data for different
$N$. 
The value $H_1$ is found by requiring that it provides the best collapse
of the curves $H_n(N)$ into one for our set of sufficiently large network sizes $N$. 
(We repeat the calculations for different $H_1$ and select the value giving the best collapse of the curves.) 
Figure~\ref{fig:ff1}(b),(d),(f) shows that for each considered case,  
such a value $H_1$ exists, and the curves $H_n$ obtained for different $N$ 
collapse into one. 
Thus, 
the ansatz of
Eq.~(\ref{ee10}) is correct, and for large $N$ the function $H_n$ is indeed independent of $N$. 
The existence of the function $H_n$ indicates the presence of the metastable localized state in the system. 




Equation~(\ref{ee10}) explains the crossover between two regimes in Fig.~\ref{fig:nVSkmax} for different system sizes and hub degrees. 
In the localized regime $\sum_n H_n/N \gg 1$, so the distribution $p_n$ is determined by the function $H_n$ (note the normalization $\sum_n p_{n,k}=1$ 
for the distribution $p_{n,k}$ of the number of infective nodes in a uniform random regular graph). 
This localized regime takes place in the region $0<\lambda \lesssim \lambda_\text{crossover}(q,N)$. On the other hand, the delocalized distribution $p_n$ coincides with $p_{n,k}$ because $\sum_n H_n/N \ll 1$ in the region  $\lambda_\text{crossover}(q,N) \lesssim \lambda<\lambda_c$. 
The crossover between the localized and delocalized states takes place in the region around $\lambda_\text{crossover}(q,N)$, where $\sum_n H_n/N \sim 1$. 
So we define $\lambda_\text{crossover}$ as the value of $\lambda$ for which $\sum_n H_n/N=1$, clearly depending on $N$ and $q$. 
One can see from Eq.~({\ref{ee10}}) that the average number of active nodes $\langle n \rangle$ in our network consists of two parts:
\begin{equation*}
\langle n \rangle= A(N)\langle n \rangle_k  + \left[1-A(N)\right] \langle n \rangle_H .
\end{equation*}
The first term is the bulk contribution,
where $\langle n \rangle_k$ is for a uniform random regular graph. The second term is the contribution of the hub, where $\langle n \rangle_H = \sum_n n H_n /\sum_n H_n$ represents the number of active nodes averaged over localized states. Both $\langle n \rangle_k$ and $\langle n \rangle_H$ are independent of $N$.
 As $N\to\infty$ the coefficient $A$ approaches $1$, and $\langle n \rangle$ approaches $\langle n \rangle_k$, on the other hand,
 for $N\ll \sum_n H_n$ the coefficient $A \ll 1$ and $\langle n \rangle \approx \langle n \rangle_H$.
%


\begin{figure}[b]
\centering
\includegraphics*[scale=0.595]{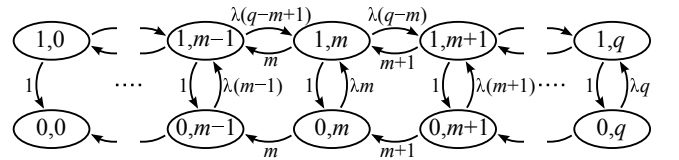}
\caption{
Graphical representation of Eq.~(\ref{ee50}) for the SIS model on a star graph.
The circles represent individual states and arrows show possible transitions between them. The number near each arrow denotes the rate of the transition.  
}
\label{f6}
\end{figure}

Let us compare the function $H_n$ with the distribution $p_n^{\text{star}}$ of the number of infective nodes in the solution of the SIS model for the star graph of the same degree as the hub. 
Estimates for the 
recovery rate of the SIS model on a star were obtained in Ref.~\cite{boguna2013nature}. We cannot use them, since for our comparison we have to describe the evolution of this system 
completely, which can be done by implementing the approach of Ref.~\cite{deroulers2004field} used for solving the SIS model on a finite fully connected graph. The state of our system at moment $t$ is 
given by the probability $P_{s,m}(t)$ that the central node is in state $s$ ($s=0,1$ is for susceptible and infective, respectively) and $m$ leaves are infective, where $0\leq m \leq q$. The evolution of this probability is exactly described by the following equations:  
\begin{eqnarray}
\partial_t P_{0,m}(t)&{=}&-(\lambda{+}1) mP_{0,m}(t){+}(m{+}1)P_{0,m{+}1}(t){+}P_{1,m}(t)
,
\nonumber
\\[5pt]
\partial_t P_{1,m}(t)&{=}&-[\lambda(q{-}m){+}m{+}1]P_{1,m}(t)+(m{+}1)P_{1,m{+}1}(t)
\nonumber
\\[5pt]
&&+\lambda(q{-}m{+}1)P_{1,m-1}(t)+\lambda mP_{0,m}(t),
\label{ee50}
\end{eqnarray}
with the boundary 
conditions $P_{0,q+1}(t) = P_{1,-1}(t) = P_{1,q+1}(t) = 0$. 
These equations 
describe the Markov chain in Fig.~\ref{f6}. 
The initial conditions are $P_{1,0}(0)=1$ and $P_{0,m}(0)=P_{1,m>0}(0)=0$, i.e., we start with only the central node infective. 
 The distribution of the number of infective nodes in a system with at least one infective node,  $p_n^{\text{star}}=[P_{1,n-1}(t)+P_{0,n}(t)]/[1-P_{0,0}(t)]$, becomes stationary for large 
 $t$. 
 This stationary distribution corresponds to the one obtained in our measurements in the quasistationary regime. 
 We obtain $p_n^{\text{star}}$ by numerically solving Eq.~(\ref{ee50}), which can be done with any desired precision for a finite $q$. 
 Figure~\ref{fig:ff1}(b),(d),(f) shows the stationary distributions $p_n^{\text{star}}$ for the stars 
 with $70$, $100$, and $150$ leaves multiplied, for the sake of comparison, by the constant $\sum_n H_n$ (thin solid  curves).
The difference between $p_n^{\text{star}}$ and $H_n$ is that in the star graph the dynamics is constrained to the hub and its nearest-neighbors, while the function $H_n$ is determined by the network activity in a wider area around the hub.

\section{Discussions and conclusions}

We have studied the 
localization of the disease in the SIS model on a random regular graph with a hub 
below the epidemic threshold. 
We found that the localized states in this system are metastable even in the infinite network limit due to the presence of the absorbing state in the SIS model. We have developed a method enabling us to quantify this phenomenon in large networks by analyzing the data of quasistationary simulations for networks of 
different sizes, specifically, the distributions of the number of infective nodes. 
We found a smooth crossover from localization on a hub at $0<\lambda\lesssim\lambda_\text{crossover}(q,N)$ to the delocalized state in the region $\lambda_\text{crossover}(q,N)\lesssim\lambda<\lambda_c$. This is quite opposite to 
the quenched mean-field theory, in which localization on a hub is predicted above a certain value of $\lambda<\lambda_c$. 
Note that for a fixed $\lambda<\lambda_c$, there is also a crossover from a delocalized to a localized state as we increase the hub degree $q$.  
We completely described the distribution of the number of infective nodes in the metastable state by the linear combination of two contributions: (i) from the uniform network and (ii) the effect of the hub, see Eq.~(\ref{ee10}). 
We have demonstrated how to extract the contribution of the hub, $H_n$,  
shown in Fig.~\ref{fig:ff1}. 
In the quasistationary state this contribution decays as $1/N$ asymptotically, which means that the effect of the hub disappears in the infinite system. 
We have compared the extracted contribution with the solution of the SIS model on a star graph having the same number of leaves as the hub in our system. 
The marked difference between $H_n$ and the distribution $p_n^{\text{star}}$ revealed  
the influence of a wide neighborhood of the hub in the infinite network. 
Notably, we considered a large network with a single hub, which could not influence the epidemic threshold $\lambda_c$. Sufficiently high concentration of strongly connected nodes (hubs) can seriously displace or even eliminate the epidemic threshold \cite{boguna2013nature,ferreira2015collective}.  

We have characterized 
the phenomenon of metastable localization of a disease  below the epidemic threshold in a 
model 
heterogeneous system and the crossover to a delocalized state. 
Our work revealed a 
drastic difference of this kind of localization from 
the standard 
one. 
Common intuition tells that a hub should be important near $\lambda_c$ where it can keep an island of disease. We find that, on the contrary, the effect of the hub actually disappears near the epidemic threshold. 
We suggest that 
our findings can be qualitatively applicable to long-lasting local 
outbreaks in  
a wide range of 
epidemic 
processes with an absorbing state on 
highly heterogeneous networks.




\begin{acknowledgments}
We thank G.~J.~Baxter, A.~V.~Goltsev, and G.~Zampieri for many stimulating discussions. 
This work was supported by the FET
proactive IP project MULTIPLEX 317532, 
and R.S.F.  
was supported by the
program Ci\^{e}ncia sem Fronteiras---CAPES, grant
No. 88881.030375/2013-01.
\end{acknowledgments}

\appendix*

\section{SIMULATION OF THE QUASISTATIONARY STATE
}

Here we outline a few issues significant for our simulations\vspace{6pt}. 


(1) 
In finite networks of the kind under consideration, finite loops are present, which contribute to the reinfection of the hub, prolonging the localized activity, see Fig.~\ref{fig:spreading}(a). This contribution disappears in the infinite size limit. To ensure that our measurements are free from the effect of finite loops, for each combination of parameters $\lambda$ and $q$ we choose a set of network sizes in the range where the lifetime of the spreading experiment is already size independent, see collapse of curves in Fig.~\ref{fig:spreading}(a). For instance, in panels (e) and (f) of Fig.~\ref{fig:ff1} we use larger system sizes than in the other panels because for $\lambda=0.2005$ and $q=150$ the effect of loops is observed for sizes  $N\sim 10^8$\vspace{6pt}.

(2) 
To access the quasistationary state we have to wait a long time after the start of the process.
It is very likely that the system falls into the absorbing state during the transient period, and we would have to restart the process many times before one realization lasts enough to reach the quasistationary state. 
Below the epidemic threshold, this is a quite inefficient approach because most of the computation time is spent simulating the transient and not the quasistationary state. 

To overcome this difficulty we use the quasistationary method \cite{ferreira2011quasistationary2,oliveira2006quasi,de2005simulate}. Within this approach, when the system falls in the absorbing state, we restart it from an active configuration taken at random from the history of active configurations visited by the process. For this, we keep a  database of a large number of active configurations, where, at random instants, we save the current configuration in a random position of the database.
As the process proceeds, the database is repeatedly updated and relaxes until the process converges to a stationary state that is independent of the initial conditions. 
In the limits of large number of states in the database and long times intervals between consecutive updates, the stationary states of the quasistationary method converges to the one of the original process. 
These two limits are important. They ensure that the configurations in the database are uncorrelated. 

The quasistationary method can be viewed as a clever way of simulating the whole ensemble of realizations of the process while collecting data from a single realization.
The advantage is that when the observed realization falls in the absorbing state we choose another at random, among the ones still active, and start collecting data from that moment forward. 
After the initial relaxation period of the history database we collect data without interruptions, dramatically reducing the computation time needed for gathering a representative statistics\vspace{6pt}.

(3) 
For each $q$ in Fig.~\ref{fig:ff1} we 
measure localized activity over $2$ orders of magnitude of system size (from $10^7$ nodes to $10^9$). 
We maximize the difference between curves for different $N$ by selecting $\lambda \approx \lambda_\text{crossover}(N{=}10^7,q)$. Recall that at $\lambda_\text{crossover}(10^7,q)$, we have $\sum_n H_n/10^7 = 1$. For $N=10^9$ the fraction of localized configurations in the history database is already small, roughly $1\% \approx \sum_n H_n/10^9$. Because of this smallness we keep a database large enough that the average number of localized configurations stored there is much larger than its fluctuations. 

For these simulations we keep a database of $10^5$ configurations.  
This number of configurations allows us to have on average about $0.01 \times 10^5=1000\gg1$ localized configurations in the networks of $10^9$ nodes.
At each time step we update a random position of the database with the current configuration with probability $0.1 dt / \langle \tau \rangle$, where $dt=1/[n(t)+\lambda l(t)]$ is the lifetime of the configuration (see main text). 
With this update probability the average time interval between consecutive updates is of roughly $10 \langle \tau \rangle$. By comparing simulations with different update intervals we found that, in all of the cases considered, the results for an average update interval of $10 \langle \tau \rangle$ are indistinguishable from those obtained with longer update intervals.
We only start to collect data after each position of the database has been updated at least $10^3$ times, allowing for the full relaxation of the history record\vspace{6pt}.

(4) 
In the quasistationary state, the rate at which the system falls in the absorbing configuration is $1/\langle \tau \rangle$. In the SIS model, this rate must be exactly equal to the probability of the active system having only one infective node, $p_1$, multiplied by the rate at which the infective node spontaneously recovers, which is $1$ in this case \cite{de2005simulate,dickman2002quasi}. 
Then to check if the resulting stationary data are correct (e.g., not spoiled by errors in the code), it is useful to compare $\langle \tau \rangle$ and $1/p_1$. These two numbers must be equal with a high precision increasing with the collected amount of data. In our simulations we verify this equality with up to 7 digits of precision. 
Note that for $\langle \tau \rangle=1/p_1$ to hold we must measure time as a continuous variable, incrementing it by $dt=1/[n(t)+\lambda l(t)]$ at each step, and define $p_n$ as the fraction of the total time that the system spends in configurations with $n$ active nodes.

\end{document}